\documentclass[3p,times,twocolumn]{elsarticle}

\usepackage{ecrc}
\volume{00}
\firstpage{1}
\journalname{Nuclear and Particle Physics Proceedings}

\runauth{A. Arbey et al.}

\jid{nppp}

\jnltitlelogo{Nuclear and Particle Physics Proceedings}

\usepackage{amssymb}

\biboptions{sort&compress}

\begin{document}

\begin{frontmatter}



\dochead{}

\title{Constraints on the CP-Violating MSSM\tnoteref{label5}}

\author[label1,label2]{A. Arbey}
\ead{alexandre.arbey@ens-lyon.fr}
\author[label3,label2]{J. Ellis}
\author[label4]{R. M. Godbole}
\author[label1,label2]{F. Mahmoudi}

\address[label1]{Univ Lyon, Univ Lyon 1, ENS de Lyon, CNRS, Centre de Recherche Astrophysique de Lyon UMR5574, F-69230 Saint-Genis-Laval, France\\[0.1cm]}
\address[label2]{Theoretical Physics Department, CERN, CH-1211 Geneva 23, Switzerland\\[0.1cm]}
\address[label3]{Theoretical Particle Physics and Cosmology Group, Department of Physics, King's College London, London WC2R 2LS, United Kingdom\\[0.1cm]}
\address[label4]{Centre for High Energy Physics, Indian Institute of Science, Bangalore, 560012, India}

\tnotetext[label5]{Based on the talk by A.A. at the Sixth Workshop on Theory, Phenomenology and Experiments in Flavour Physics, Capri, June 2016.} 

\begin{abstract}
We discuss the prospects for observing CP violation in the MSSM with six CP-violating phases, using a geometric approach to maximise CP-violating observables subject to the experimental upper bounds on electric dipole moments. We consider constraints from Higgs physics, flavour physics, the dark matter relic density and spin-independent scattering cross section with matter.

\end{abstract}


\end{frontmatter}

\section{Introduction}

In the minimal supersymmetric extension of the Standard Model (MSSM), there are possibilities of having new sources of CP violation beyond the Cabibbo Kobayashi Maskawa (CKM) phase. However, experimental constraints in particular from the measurements of electric dipole moments (EDMs, see Tab.~\ref{tab:EDMs}) limit strongly the values of the additional phases. Here we consider the maximally CP-violating, minimally flavour-violating (MCPMFV) model that contains 6 new CP-violating phases: 3 phases $\Phi_{1,2,3}$ in the masses of the U(1), SU(2) and SU(3) gauginos, and 3 phases $\Phi_{t,b,\tau}$ in the trilinear soft supersymmetry breaking couplings $A_{t,b,\tau}$ of the third-generation stop, sbottom and stau sfermions, respectively. Brute force to sample randomly the values of the phases imposing all the experimental constraints (from Higgs, flavour physics, dark matter and EDMs) appears to be inefficient and one would need an optimised procedure such as a geometric approach~\cite{Ellis:2010xm,Arbey:2014msa}. We study signatures of CP violation considering the lightest neutral Higgs boson to be the observed one and the lightest supersymmetric particle to be the lightest neutralino, and study in particular the CP asymmetry in $b \to s \gamma$, $B_s$ meson mixing $\Delta M_{B_s}$, and CP-violating couplings of the heavier neutral Higgs bosons in several representative MSSM scenarios. Extended discussions can be found in \cite{Arbey:2014msa}.

\section{Methodology}

\begin{table}[!t]
\begin{center}
{\renewcommand*{\arraystretch}{1.4}
\begin{tabular}{|c|c|c|}
 \hline
 EDM & Upper limit (e.cm) & Reference\\
 \hline\hline
 Thallium & $1.3\times10^{-24}$ & \cite{Regan:2002ta}\\
 \hline
 Mercury & $3.5\times10^{-29}$ & \cite{Griffith:2009zz}\\
 \hline
 Neutron & $4.7\times10^{-26}$ &  \cite{Baker:2006ts}\\
  \hline
 Thorium monoxide & $1.1\times10^{-28}$ & \cite{Baron:2013eja} \\
 \hline
\end{tabular}%
}
\caption{\it 95\% CL upper limits on the EDMs used in this study.\label{tab:EDMs}}
\end{center}
\end{table}

We consider here two different MSSM scenarios: the CPV-CMSSM and the CPV-pMSSM, which are extensions of the standard CMSSM and pMSSM scenarios incorporating six additional phases $\Phi_{1,2,3,t,b,\tau}$ to account for CP violation. We perform flat random scans on the standard parameters, similarly to \cite{Arbey:2011un,Arbey:2011aa,Arbey:2012bp}. The phases however are very severely constrained by the electric dipole moment (EDM) measurements given in Tab.~\ref{tab:EDMs}, so that the use of an optimised geometric technique becomes necessary. The muon EDM has not been used in our analysis since its experimental upper bound \cite{Bennett:2008dy} only provides very weak constraints that are not competitive with the constraints from the other EDMs.

Let us consider the four EDMs, $E^{a,b,c,d}$, of Tab.~\ref{tab:EDMs} in the small phase approximation, with
\begin{equation}
E^i \; \simeq \; {\mathbf \Phi}.{\mathbf E}^i \, ,
\end{equation}
where ${\bf \Phi} \equiv \Phi_\alpha = \Phi_{1,2,3,t,b,\tau}$ and ${\mathbf E}^i \equiv \partial E^i/\partial {\mathbf \Phi}$,
and an additional CP-violating observable $O$ in the the small phase approximation, such that
\begin{equation}
 {\mathbf O} \equiv \partial O/\partial {\mathbf \Phi} \,.
\end{equation}
The optimal direction that maximises the observable $O$ and is orthogonal to the EDM vectors $E^{a,b,c,d}_\alpha$ is given by
\begin{equation}
 \hspace*{-0.5cm}\Phi_\alpha = \epsilon_{\alpha\beta\gamma\delta\mu\eta} \, \epsilon_{\eta\nu\lambda\rho\sigma\tau} \, E^a_\beta \, E^b_\gamma \, E^c_\delta \, E^d_\mu \, O_\nu \, E^a_\lambda \, E^b_\rho \, E^c_\sigma \, E^d_\tau \, ,
 \label{eq:geometric}
\end{equation}
with an unknown normalisation factor.

For each choice of the CP-conserving parameters, we fix the phases to $0^\circ$ or $\pm180^\circ$, and compute the optimal direction using Eq.~(\ref{eq:geometric}). Sets of phases are chosen randomly along this direction, then moved by $20^\circ$ along the favoured direction. Subsequently, the favoured direction is recomputed at this new position. We iterate this procedure up to $100^\circ$.

The SUSY mass spectra and couplings, and EDM constraints are computed with {\tt CPsuperH}~\cite{Lee:2003nta,Lee:2007gn,Lee:2012wa}. The thorium monoxide EDM is computed following \cite{Cheung:2014oaa}. Flavour observables are calculated with {\tt SuperIso}~\cite{Mahmoudi:2007vz,Mahmoudi:2008tp} and {\tt CPsuperH}. The dark matter relic density is computed with {\tt SuperIso Relic}~\cite{Arbey:2009gu} and {\tt micrOMEGAs}~\cite{Belanger:2013oya}, and the later is also used to compute dark matter direct detection observables. Finally, {\tt HiggsBounds}~\cite{Bechtle:2013wla} is used to impose the Higgs constraints. In contrast to the CP-conserving MSSM scenarios, here the couplings and mixing matrices can be complex, and the three Higgs bosons can also be mixed and have scalar and pseudoscalar components, leading to three states $h_{1,2,3}$, named in the order of increasing masses.

\section{Constraints in the CMSSM}

\begin{figure}[t]
\begin{center}
\includegraphics[width=0.5\columnwidth]{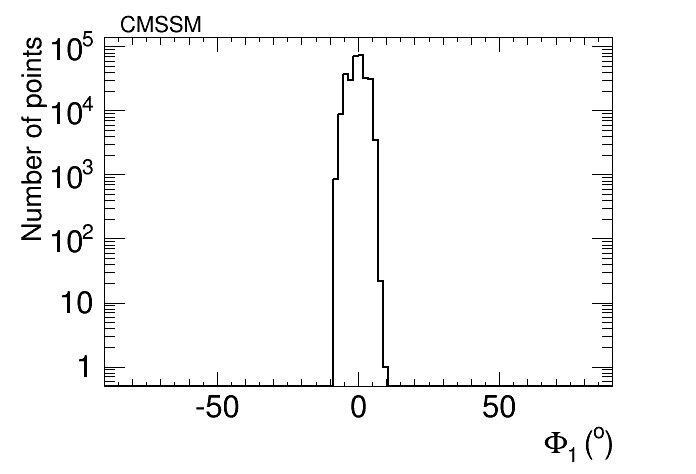}\includegraphics[width=0.5\columnwidth]{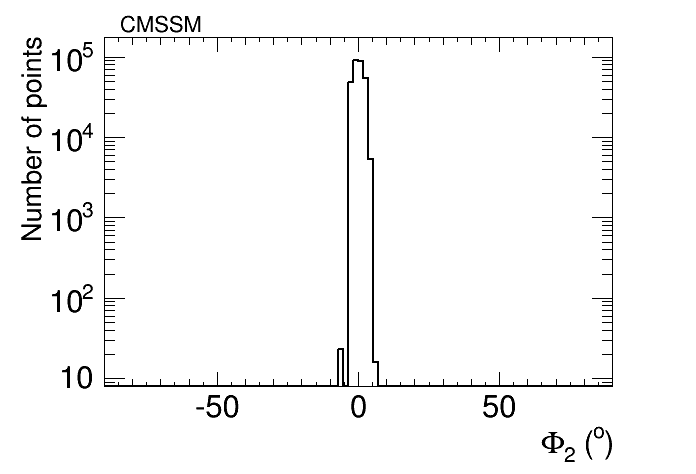}\\
\includegraphics[width=0.5\columnwidth]{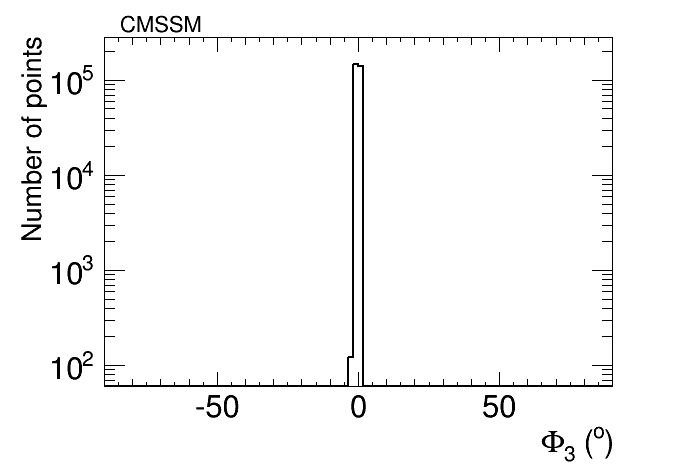}\includegraphics[width=0.5\columnwidth]{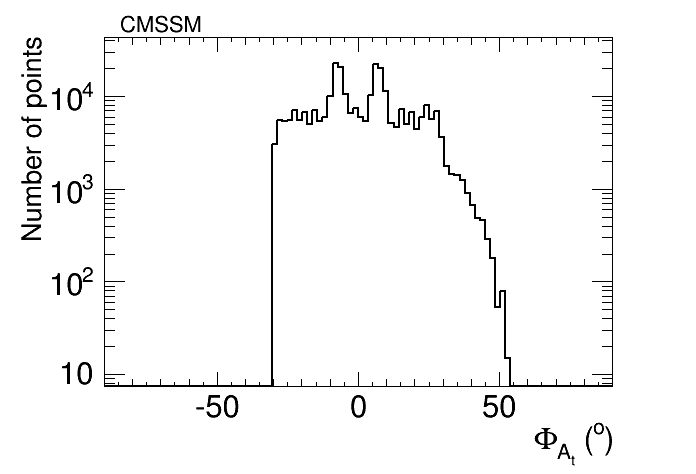}\\
\includegraphics[width=0.5\columnwidth]{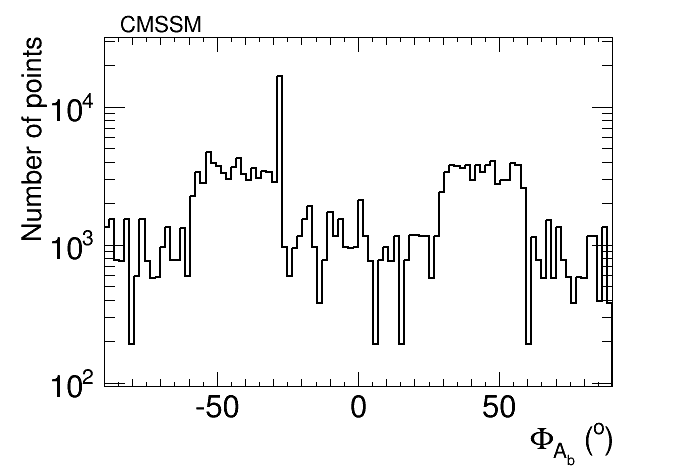}\includegraphics[width=0.5\columnwidth]{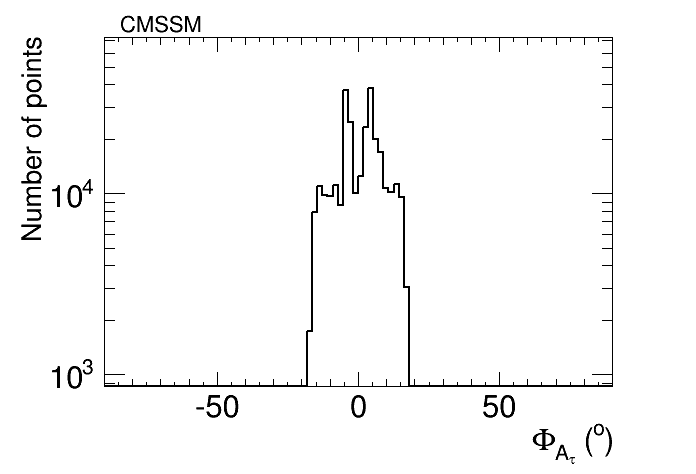}\\[0.3cm]
\caption{Number of CMSSM points allowed by the EDM constraints, starting from a flat distribution, as a function of the phases.\label{fig:cmssm-phases}}
\end{center}
\end{figure}

We consider the CMSSM as a benchmark scenario to study the effects of the CP phases. For this purpose, we focus on the CP-conserving CMSSM best-fit point found in a global analysis~\cite{Buchmueller:2013rsa}:
\begin{center}
$m_0=670$ GeV, $m_{1/2}=1040$ GeV, $A_0=3440$ GeV, $\tan\beta=21$\, ,
\end{center}
and vary the phases subsequently. To test the efficiency of the geometric approach, we vary the phases randomly, then employ the geometric approach and finally apply the constraints. 600.000 points are generated for each case. Imposing the constraints removes about 85\% of the points in the purely random approach, while only 50\% of the points are removed with the geometric approach, showing the efficiency of the latter. 

Fig.~\ref{fig:cmssm-phases} shows the distribution of the phases obtained by imposing the EDM constraints. Clearly, the phases $\Phi_1$, $\Phi_2$, $\Phi_3$ are strongly constrained. 
\begin{figure}[t]
\begin{center}
\includegraphics[width=0.5\columnwidth]{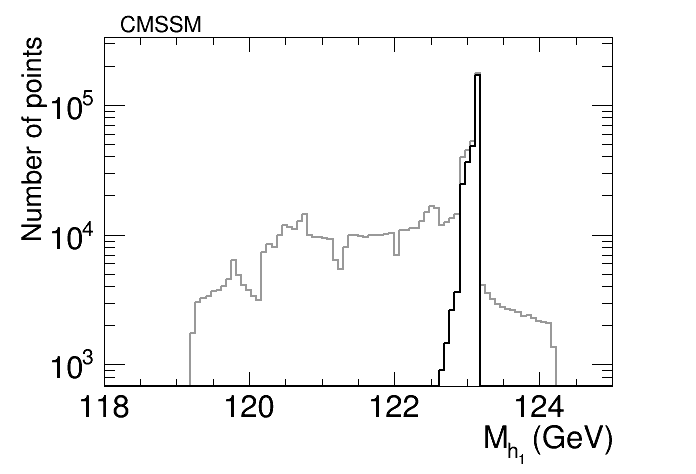}\includegraphics[width=0.5\columnwidth]{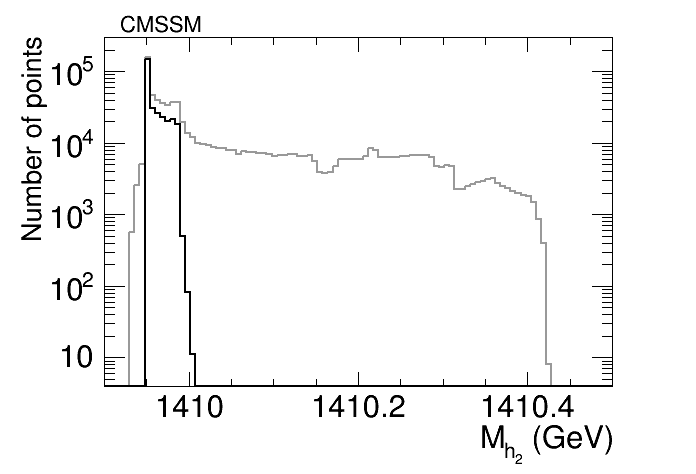}\\
\includegraphics[width=0.5\columnwidth]{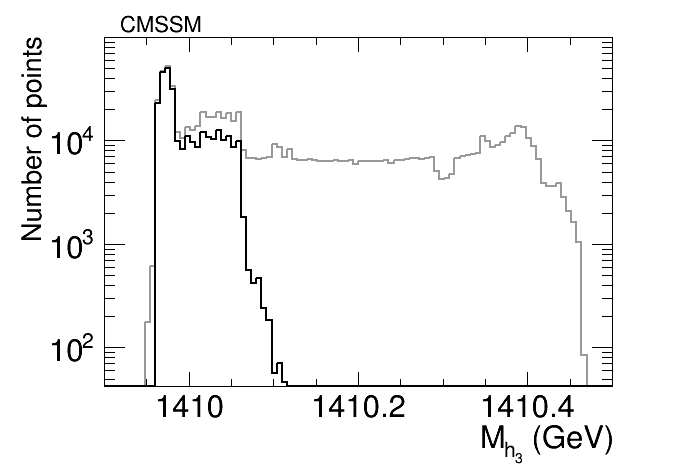}
\caption{Distribution of the Higgs masses for the considered CMSSM scenario. The gray line corresponds to the number of points before applying the EDM constraints, and the black one after imposing them.\label{fig:cmssm-higgses}}
\end{center}
%
\begin{center}
\includegraphics[width=0.5\columnwidth]{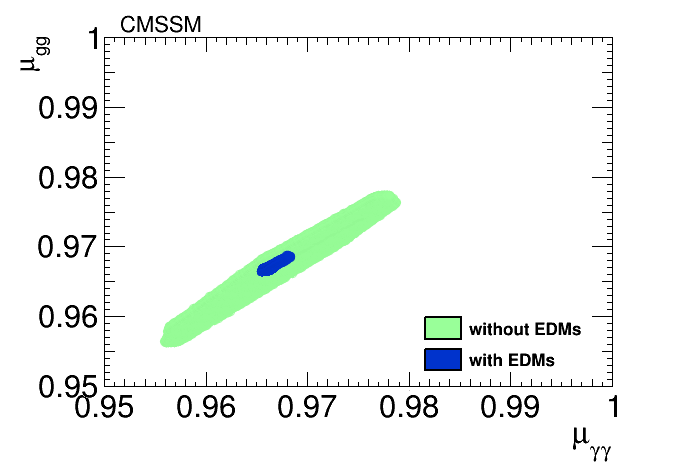}\includegraphics[width=0.5\columnwidth]{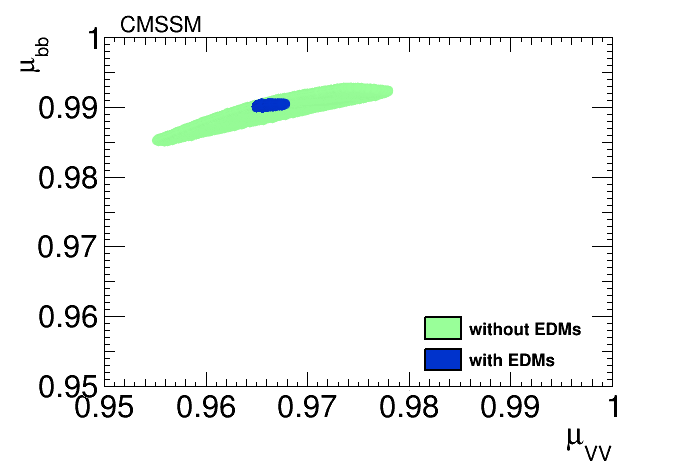}
\caption{Lighter Higgs signal strengths in $(\mu_{\gamma\gamma},\mu_{gg})$ (left panel) and $(\mu_{VV},\mu_{bb})$ (right panel) planes before and after imposing the EDM constraints.\label{fig:cmssm-higgs2}}
\end{center}
\end{figure}
The EDM constraints also severely restrict the possible Higgs masses, as shown in Fig.~\ref{fig:cmssm-higgses}.
In addition, the EDM constraints impose the Higgs signal strengths to be very close to 1, as demonstrated in Fig.~\ref{fig:cmssm-higgs2}.

More generically, when varying all the CMSSM parameters, the EDM measurements strongly limit the CP-violating CMSSM to be very close to the CP-conserving CMSSM (see \cite{Arbey:2014msa} for more details). This conclusion however does not hold in an unconstrained scenario such as the phenomenological MSSM (pMSSM).

\section{Constraints in the pMSSM}

\begin{figure}[t]
\begin{center}
\includegraphics[width=0.5\columnwidth]{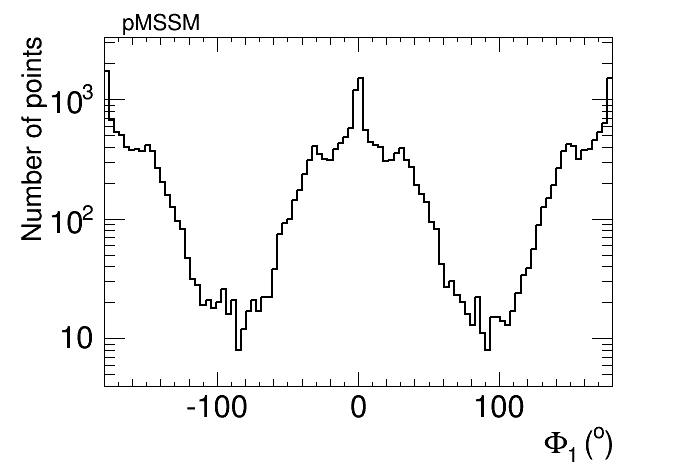}\includegraphics[width=0.5\columnwidth]{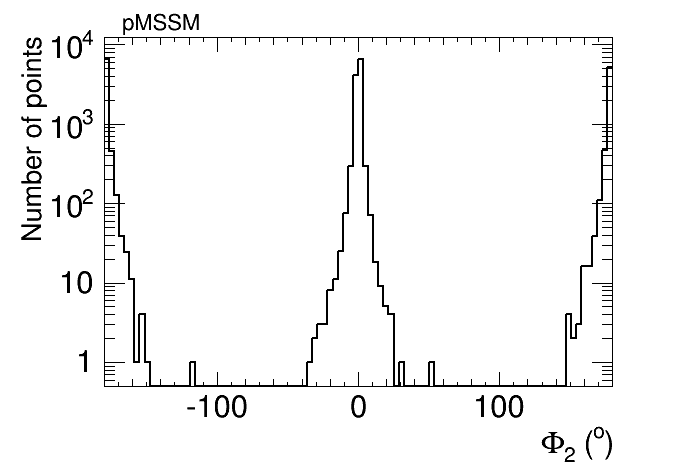}\\
\includegraphics[width=0.5\columnwidth]{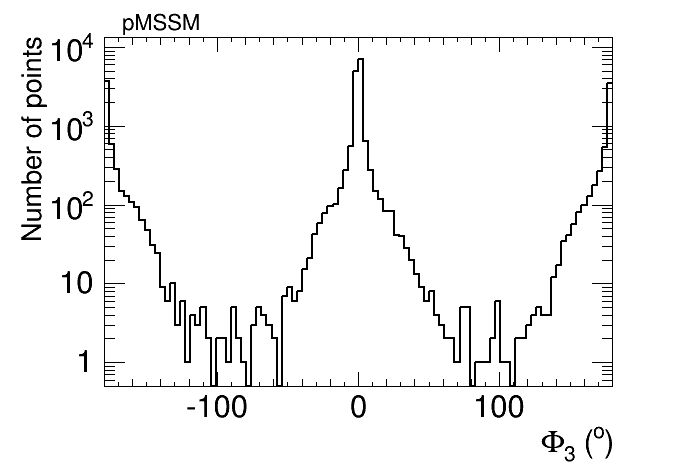}\includegraphics[width=0.5\columnwidth]{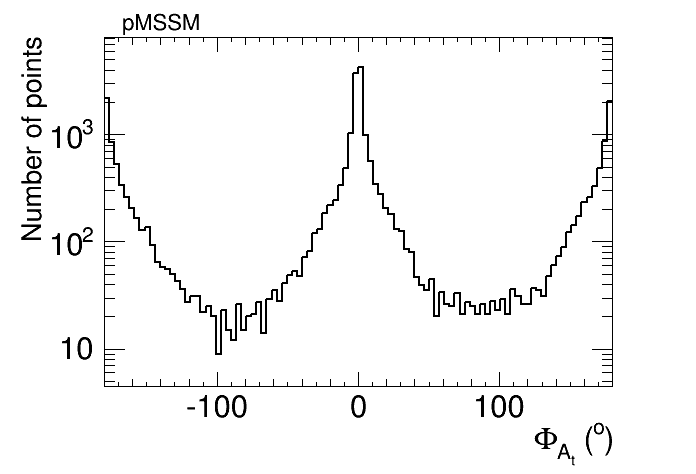}\\
\includegraphics[width=0.5\columnwidth]{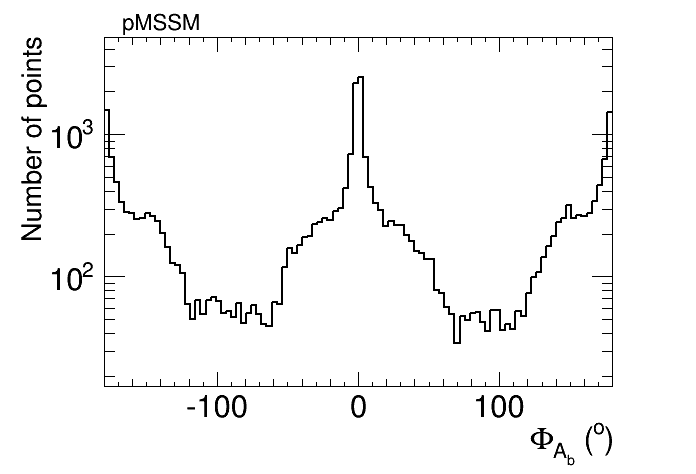}\includegraphics[width=0.5\columnwidth]{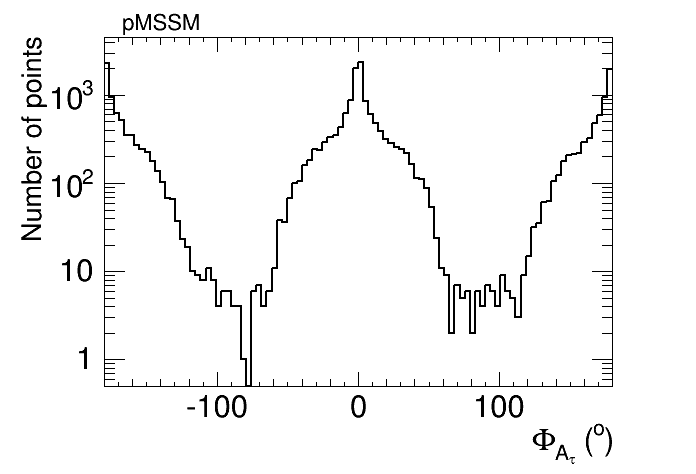}
\caption{Number of pMSSM points allowed by the EDM constraints as a function of the phases.\label{fig:pmssm-phases}}
\end{center}
%
\begin{center}
\includegraphics[width=0.5\columnwidth]{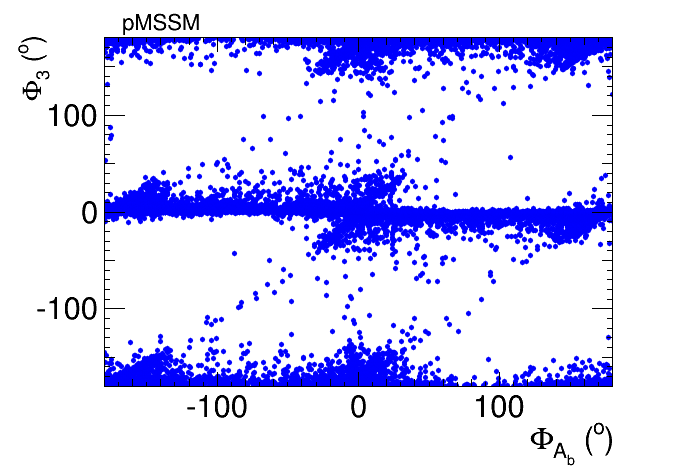}\includegraphics[width=0.5\columnwidth]{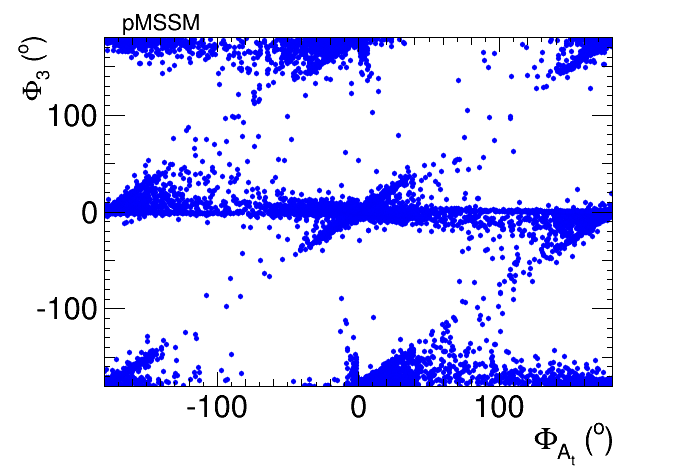}
\caption{pMSSM points surviving the EDM constraints in the $(\Phi_b,\Phi_3)$ (left panel) and $(\Phi_t,\Phi_3)$ (right panel) parameter planes.\label{fig:pmssm-phases2}}
\end{center}
\end{figure}

To study the pMSSM, we first vary the 19 CP-conserving pMSSM parameters. We then use the geometric approach to choose points with phases between $-180^\circ$ and $+180^\circ$. 40 million such points have been generated with gluino and squark masses compatible with the LHC limits. We then impose the $h_1$ mass to lie between 121 and 129 GeV and the lightest neutralino to be the lightest supersymmetric particle so that it constitutes a dark matter candidate. This reduces the number of points to about one million. After imposing the EDM constraints, we are left with 15.000 points. The distribution of the CP phases is shown in Fig.~\ref{fig:pmssm-phases}.

As can be seen, the $\Phi_2$ phase is particularly constrained, and the other phases can take any values. In Fig.~\ref{fig:pmssm-phases2}, the phases of the points compatible with the EDM constraints are drawn in 2 dimensional plots. They reveal that the EDM constraints impose correlations between the phases related to the gluino and third-generation squarks, with prefered directions that are revealed by the geometric approach.

In the following, we study the influence of CP violation on the dark matter, Higgs and flavour sectors.

\subsection{Dark matter sector}

\begin{figure}[t]
\begin{center}
\includegraphics[width=0.8\columnwidth]{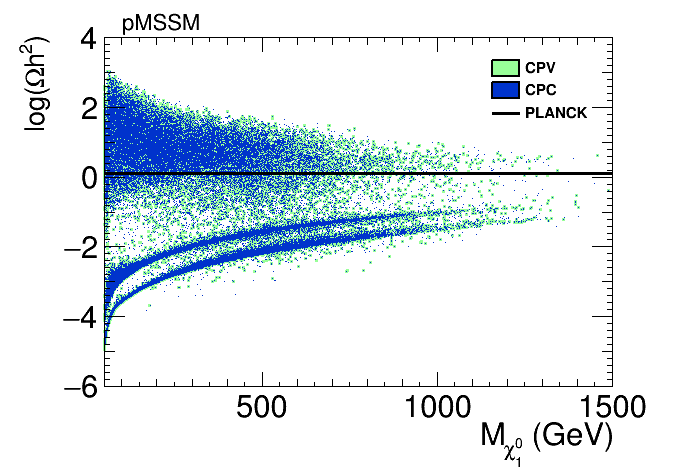}\\
\includegraphics[width=0.8\columnwidth]{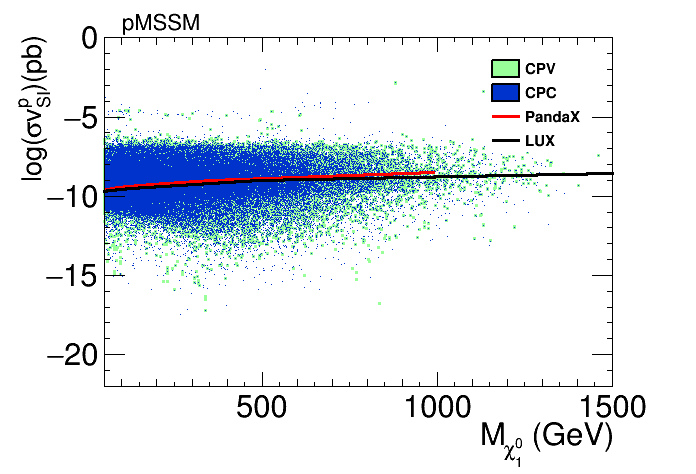}
\caption{pMSSM points compatible with the EDM constraints in the relic density vs. neutralino 1 mass (upper panel) and neutralino spin-independent scattering cross section with proton vs. neutralino 1 mass (lower panel). The blue points are the points without CP violation, and the green points have CP-violating phases. The black line on the upper plot corresponds to the central value of the dark matter density measured with cosmological observations, and the red and black curves on the lower plot to the PandaX and LUX 2016 upper limits on the scattering cross section.\label{fig:pmssm-dm}}
\end{center}
\end{figure}

We first compute the neutralino relic density and compare the results to the cosmological cold dark matter density $\Omega_c h^2 \sim 0.11$ \cite{Ade:2015xua}. The results are presented in the upper panel of Fig.~\ref{fig:pmssm-dm}. We notice that the points with CP conservation and the ones with CP violation are similarly spread, such that the CP phases do not add much content to the picture. The points above the cosmological measurements have mostly bino-like neutralino, and the two strips below the dark matter line correspond to wino and higgsino-like neutralinos. The points above the line can be considered as excluded, since they lead to an overdensity of dark matter, provided the pre-Big Bang nucleosynthesis period is dominated by radiation \cite{Arbey:2008kv,Arbey:2009gt}, so that most of the bino-like neutralinos are excluded. CP violation allows for more such points to survive, without changing the picture. For the points below the line the neutralino relic density could only partially account for the whole dark matter density, and they can therefore be considered as still allowed.

We also consider dark matter direct detection and the latest upper limits provided by the PandaX \cite{Tan:2016zwf} and LUX \cite{Akerib:2016vxi} experiments. The results are shown in the lower panel of Fig.~\ref{fig:pmssm-dm}. Similarly to the case of the relic density, we see that while CP violation allows for some more spread of the points in comparison to the CP-conserving case, it does not add much to the picture. This shows that dark matter observables are unlikely to disentangle CP-violating scenarios form the CP-conserving ones.

\subsection{Higgs sector}

\begin{figure}[t]
\begin{center}
\includegraphics[width=0.8\columnwidth]{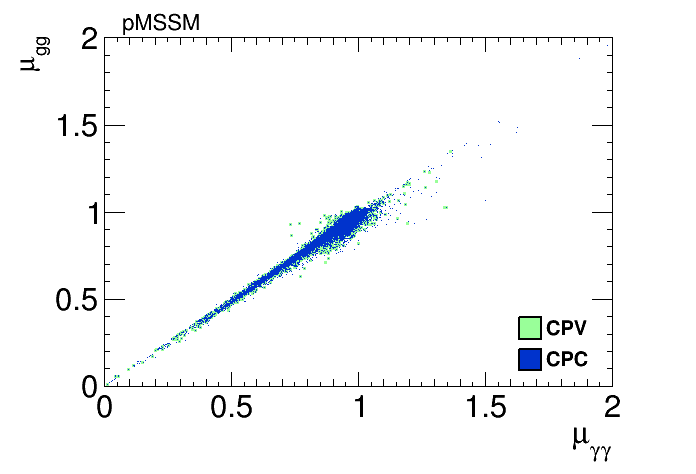}\\
\includegraphics[width=0.8\columnwidth]{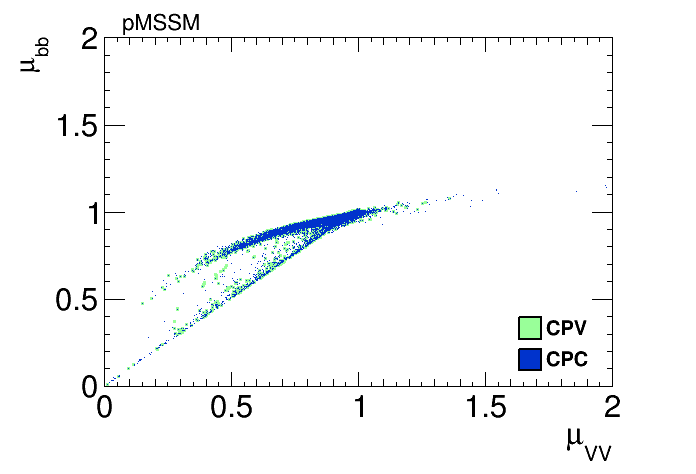}
\caption{pMSSM points compatible with the EDM constraints in the $\mu_{gg}$ vs. $\mu_{\gamma\gamma}$ (upper panel) and  $\mu_{bb}$ vs. $\mu_{VV}$ (lower panel) parameter planes. The blue points are the points without CP violation, and the green points have CP-violating phases.\label{fig:pmssm-higgs}}
\end{center}
\end{figure}

First, we assume the lightest Higgs boson to be the one discovered at the LHC by imposing its mass to lie between 121 and 129 GeV, so as to allow for theoretical uncertainties in its calculation, and we study the modifications to the signal strengths due to CP violation. The signal strength in the decay channel $h_1 \to XX$ is defined as the ratio of the cross section production of the $h_1$ times its decay branching ratio to $XX$ over the Standard Model values. The LHC results for the measured decay modes show that the signal strengths are compatible with 1 \cite{Khachatryan:2016vau}. Our results are summarised in Fig.~\ref{fig:pmssm-higgs}. As can be seen, the set of points with CP violation are not distinct from the one with CP conservation, so that the signal strength measurement will not allow to probe the difference between the two cases.

\begin{figure}[t]
\begin{center}
\includegraphics[width=0.5\columnwidth]{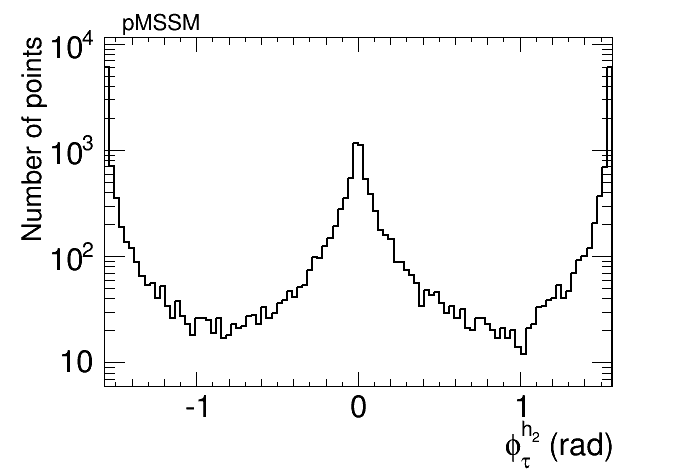}\includegraphics[width=0.5\columnwidth]{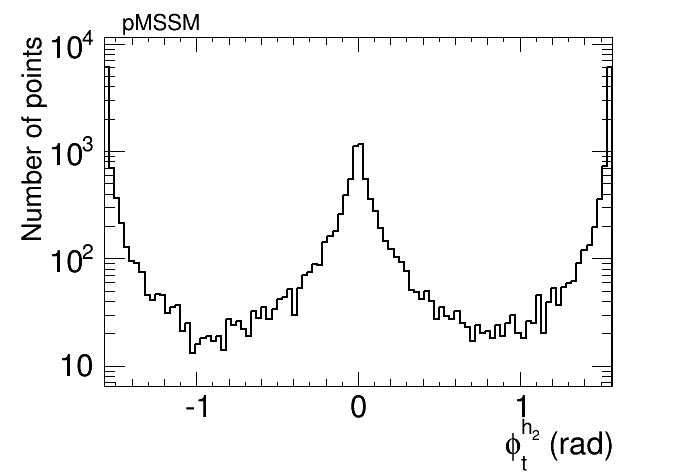}
\caption{Distribution of the $\phi^{h_2}_\tau$ and $\phi^{h_2}_t$ angles in the CPV-pMSSM after imposing the EDM constraints.\label{fig:pmssm-higgs2}}
\end{center}
\end{figure}

The Higgs sector however provides other observables which can allow to probe CP violation. The Higgs bosons can be admixtures of scalar and pseudoscalar components. On the one side, the measurements of the $h_1$ properties showed that it is mostly a scalar, and the pseudoscalar component could only be tiny, and adding the EDM constraints limits the pseudoscalar component to be negligible. On the other side, the two heavier Higgses can be mixtures of scalar and pseudoscalar components. We denote $g_{S,P}^{h_i {\bar f} f}$ the scalar and pseudoscalar couplings of $h_i$ to ${\bar f} f$ and define 
\begin{equation}
\tan \phi^{h_i}_\tau \; \equiv \; \frac{g_P^{h_i \tau \tau}}{g_S^{h_i \tau \tau}}\;, \quad
\tan \phi^{h_i}_t \; \equiv \; \frac{g_P^{h_i {\bar t} t}}{g_S^{h_i {\bar t} t}}\;.
\end{equation}
The angles $\phi^{h_i}_\tau$ and $\phi^{h_i}_t$ characterise the scalar and pseudoscalar couplings to the $\tau$ leptons and top quarks, respectively. An angle close to 0 means a mostly scalar coupling, while $\pi/2$ means a pure pseudoscalar coupling. In Fig.~\ref{fig:pmssm-higgs2}, we show the distributions of the angles of the $h_2$ after imposing the EDM constraints. Similar plots can be obtained for $h_3$ \cite{Arbey:2014msa}. These results demonstrate that strongly-mixed states are still allowed for the heavy Higgses, and that a measurement of the $\tau\tau$ and $t\bar{t}$ spin correlations in $h_{2,3}$ decays may reveal the CP-violating nature of the MSSM.

\subsection{Flavour sector}

\begin{figure}[t]
\begin{center}
\includegraphics[width=0.5\columnwidth]{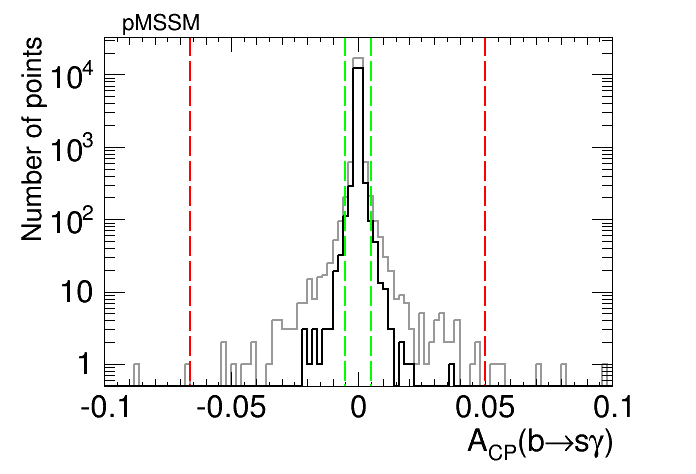}\includegraphics[width=0.5\columnwidth]{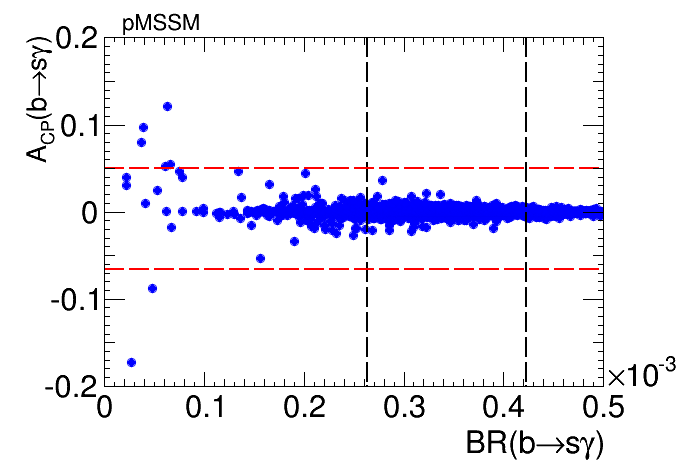}
\caption{In the left panel, the distributions of the values of the CP asymmetry of $B \to X_s \gamma$ for the pMSSM points are displayed. The gray curve corresponds to the points before the EDM constraints, and the black curve after the EDM constraints. The red dashed lines correspond to the current limits and the green dashed lines to the prospective Belle-II limits. In the right panel, the pMSSM points passing the EDM constraints in the CP asymmetry vs. branching ratio of $B \to X_s \gamma$ parameter plane are shown. The lines correspond to the current limits. \label{fig:pmssm-bsg}}
\end{center}
%
\begin{center}
\includegraphics[width=0.8\columnwidth]{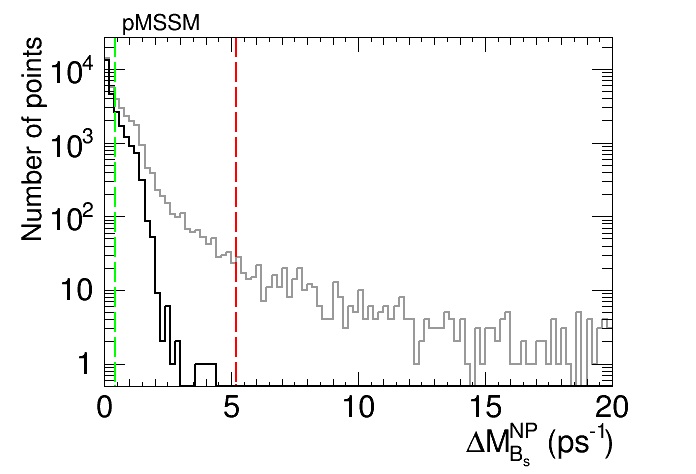}
\caption{Distribution of the values of $\Delta M^{NP}_{B_s}$ for the pMSSM points. The gray curve corresponds to the points before the EDM constraints, and the black curve after the EDM constraints. The red dashed line shows the current limit and the green dashed line the prospective limit after a factor 10 improvement in the form factor determination.\label{fig:pmssm-DeltaMB}}
\end{center}
\end{figure}

CP violation is intimately related to flavour physics, in particular because of the CP-violating phase present in the CKM matrix. Flavour-changing neutral currents constitute important observables to probe for new physics and CP violation. In particular, the inclusive decay $B \to X_s \gamma$ measured by $B$-factories provides valuable constraints. In Fig.~\ref{fig:pmssm-bsg}, we present the constraints imposed by the measurements of the branching ratio and CP asymmetry of this decay on our pMSSM points passing the EDM constraints. For comparison, we show the current \cite{Olive:2016xmw} and prospective \cite{Aushev:2010bq} limits. As can be seen, the current limits are superseded by the EDM constraints, but the perspectives to discover CP violation in the MSSM at Belle-II are promising.

Another important flavour observable is the $B_s$ meson mixing $\Delta M_{B_s}$. 
As can be seen in Fig.~\ref{fig:pmssm-DeltaMB}, the CP-violating MSSM contributions to $\Delta M^{NP}_{B_s}$ are in general below the present upper limit, which is dominated by theoretical uncertainties. If these could be
reduced, $\Delta M_{B_s}$ could also provide an interesting and complementary constraint on the phases, enabling them to be determined experimentally, in principle.

\section{Conclusion}

We have explored the effects of CP violation in the MSSM using an iterative geometrical approach within the
maximally CP-violating, minimal flavour-violating framework with 6 CP-violating phases, applying constraints from the Higgs mass and signal strengths, flavour physics, dark matter relic density and scattering cross section with matter.

In the CMSSM scenario, we found relatively little scope for large deviations from the CP-conserving case, e.g., in the masses of the Higgs bosons and the spin-independent dark matter scattering cross section. Moreover, we found that only very small values of $A_{CP}$ would be possible in this case, and the new physics contribution to $B_s$ meson mixing would not be observable.

In the pMSSM scenario, the $A_{CP}$ could be as large as $\sim 3$\%, within the reach of the Belle-II experiment. We found in this scenario that $\Delta M^{NP}_{B_s}$ could be large enough to be observable with a prospective reduction in the theoretical uncertainty in the Standard Model calculation of $B_s$ mixing.

The CP-violating phases in the $h_1 \tau^+ \tau^-$ and $h_1 {\bar t} t$ couplings are in general small in the MSSM. However, the phases in the $h_{2,3} \tau^+ \tau^-$ and $h_{2,3} {\bar t} t$ couplings can be quite large, and may present interesting prospects for future experiments.

In general, we showed that the EDM constraints do not force all the six CP-violating phases to be small, and in some variants of the MSSM there could be observable signatures of CP violation beyond the Standard Model, such as $A_{CP}$ in the $B \to X_s \gamma$ decay and $\Delta M^{NP}_{B_s}$.

\vspace*{-0.4cm}

\section*{Acknowledgements}
\vspace*{-0.2cm}
A.A. and F.M. would like to thank the organisers for their invitation and the fruitful workshop, and would like to express a special thank to the Mainz Institute for Theoretical Physics (MITP), the Universit\`a di Napoli Federico II and INFN for their hospitality and support. 
\vspace*{-0.5cm}

\nocite{*}
\bibliographystyle{elsarticle-num}
\bibliography{capri2016-arbey}


\end{document}